\documentclass[12pt]{article}
 
\usepackage[english]{babel}
\usepackage{latexsym}
\usepackage{epsfig}
 
\begin{document}
 
\title{A perturbative study of delocalisation transition in
one-dimensional models with long-range correlated disorder}

\author{L.~Tessieri \\
{\it Department of Chemistry, Simon Fraser University,} \\
{\it Burnaby, British Columbia, Canada V5A 1S6} }
 
\date{10th June 2002}
 
\maketitle
 
\begin{abstract}
We study the delocalisation transition which takes places in
one-dimensional disordered systems when the random potential
exhibits specific long-range correlations. We consider the
case of weak disorder; using a systematic perturbative approach,
we show how the delocalisation transition brings about a change of
the scaling law of the inverse localisation length which ceases to
be a quadratic function of the disorder strength and assumes a
quartic form when the threshold separating the localised phase from
the extended one is crossed.
\end{abstract}
 
{PACS numbers: 73.20.Jc, 72.15.Rn, 72.10.Bg}

Recently, one-dimensional (1D) models with correlated disorder have
become the object of intense scrutiny, as a consequence of the
discovery that specific long-range correlations of the random potential
can create a continuum of extended states and give rise to mobility
edges~\cite{Mou98,Izr99}.
The emergence of a phase of extended states implies that, when the
disorder exhibits appropriate long-range spatial correlations,
even 1D models can display a metal-insulator transition analogous
to the Anderson transition which takes place in three-dimensional
disordered systems.
Although the results of the works~\cite{Mou98,Izr99} were obtained
for discrete lattices, it was soon found that long-range correlations
can produce analogous phenomena also in continuous disordered
models~\cite{Izr01}.
The mobility edge predicted in~\cite{Izr99} received an experimental
confirmation in~\cite{Kuh00}, where the phenomenon was observed in
the transmission of microwaves in a single-mode guide with a random
array of correlated scatterers.
Further progress along this line of research would make possible the
realization of 1D devices with specified mobility edges which could
be used as window filters in electronic, acoustic or photonic structures,
as suggested in~\cite{Iku01}.

The theoretical and practical importance of the delocalisation
transition in 1D models makes desirable to obtain a complete
understanding of the role played by correlations in the formation
of mobility edges.
Unfortunately, the theoretical comprehension of the phenomenon
achieved in~\cite{Izr99} and in the successive analytical works
is not totally satisfactory, because it rests on the assumption of
weak disorder and on the use of a second-order approximation
scheme. Thus, the delocalisation transition has been defined by the
condition that the second-order term of the inverse localisation length
(or Lyapunov exponent) vanish over a continuum interval of electronic
energies, while higher-order terms of the Lyapunov exponent have been
completely neglected.
This simplifying approach is effective but incomplete, because it does
not clarify the true nature of the ``extended'' states, leaving open
the question of whether they are completely delocalised or not.
The aim of this work is to shed light on this point and thus provide
a better understanding of the delocalisation transition in 1D models.
To this end, we make use of a systematic perturbative approach that
allows us to compute the fourth-order term of the Lyapunov exponent.
Armed with this result, we are able to ascertain the nature of the
extended states and to discuss the differences between Gaussian and
non-Gaussian potentials which appear in the localisation properties
of 1D models at this refined level of description.

In what follows, we consider the Hamiltonian continuous model defined
by the Schr\"{o}dinger equation
\begin{equation}
-(\hbar^{2}/2m) \psi''(x) + \varepsilon U(x) \psi(x) =
E  \psi(x)
\label{mod}
\end{equation}
which describes an electron of energy $E$ moving under the
influence of a random potential $U(x)$. We study the case of weak
disorder, identified by the condition $\varepsilon \rightarrow 0$.
To simplify the form of the equations, in the rest of this
paper we adopt a system of units such that $\hbar^{2}/2m = 1$.
We consider electrons of positive energy, so that we can write
$E = k^{2}$; since the wavevector $k$ enters model~(\ref{mod}) only
via the energy, we can restrict our attention to the case $k>0$ without
loss of generality.

We define the statistical properties of the disorder through the
moments of the potential $U(x)$. We assume that the model under study is
spatially homogeneous in the mean; as a consequence, the $n$-th moment of
the potential
\begin{displaymath}
\chi_{n} (x_{1},x_{2},\ldots,x_{n-1}) =
\langle U(x) U(x+x_{1}) \ldots U(x+x_{n-1}) \rangle
\end{displaymath}
depends only on $n-1$ relative coordinates. Here and in the
following we use angular brackets $\langle \ldots \rangle$ to
denote the average over disorder realizations.
We also want the average features of model~(\ref{mod}) to be invariant
under a change of sign of the potential; therefore we assume that all
moments of odd order vanish.
Finally, for the sake of mathematical convenience, we temporarily suppose
that a correlation length $l_{c}$ can be defined such that the values
$U(x_{1})$ and $U(x_{2})$ of the potential are statistically
independent when the distance between the points $x_{1}$ and $x_{2}$
is much larger than $l_{c}$. To study the case of long-range correlations,
we will subsequently relax this hypothesis by taking the limit
$l_{c} \rightarrow \infty$.

To determine the average behaviour of the eigenstates of Eq.~(\ref{mod}),
we introduce the inverse localisation length
\begin{equation}
\lambda (k)  = \lim_{x \rightarrow \infty} \frac{1}{4x}
\ln \langle \psi^{2}(x) k^{2} + \psi'^{2}(x) \rangle .
\label{loclen}
\end{equation}
As a first step to compute the Lyapunov exponent~(\ref{loclen}), we derive
a differential equation for $\psi^{2}(x)$ and $\psi'^{2}(x)$. Using
Eq.~(\ref{mod}) as a starting point, it is easy to show that the vector
\begin{displaymath}
u(x) = \left( \begin{array}{c} \psi^{2}(x) \\
                               \psi'^{2} (x)/k^{2} \\
                               \psi(x) \psi'(x)/k
              \end{array} \right)
\end{displaymath}
obeys the equation
\begin{equation}
\frac{d u}{dx} =
\left( {\bf{A}} + \varepsilon \xi(x) {\bf{B}} \right) u ,
\label{stoceq}
\end{equation}
where $\xi(x) = U(x)/k^{2}$ is the scaled potential and the symbols
$\bf{A}$ and $\bf{B}$ stand for the matrices
\begin{displaymath}
\begin{array}{ccc}
\bf{A} = \left( \begin{array}{ccc} 0 &  0 &  2k \\
                                   0 &  0 & -2k \\
                                  -k &  k &  0
                \end{array}
         \right) &
\mbox{and} &
\bf{B} = \left( \begin{array}{ccc} 0 & 0 &  0 \\
                                   0 & 0 & -2k \\
                                  -k & 0 &  0
                \end{array}
         \right) .
\end{array}
\end{displaymath}
At this point our problem consists in extracting information on the
average vector $\langle u(x) \rangle$ from the stochastic equation for
$u(x)$; this can be done using a method developed by Van Kampen~\cite{Kam74},
which replaces Eq.~(\ref{stoceq}) with an ordinary differential equation
of the form
\begin{equation}
\frac{d \langle u(x) \rangle}{dx} = {\bf{K}}(x) \langle u(x) \rangle .
\label{nonstoc}
\end{equation}
The generator ${\bf{K}}(x)$ in Eq.~(\ref{nonstoc}) is a sure operator
that can be expressed in terms of a cumulant expansion
\begin{equation}
{\bf{K}}(x) = \sum_{n=0}^{\infty} \varepsilon^{n} {\bf{K}}_{n}(x)
\label{cumexp}
\end{equation}
where the partial generators ${\bf{K}}_{n}(x)$ are functions of specific
combinations (known as ``ordered cumulants'') of the moments of the
random matrix
\begin{equation}
{\bf{M}}(x) = \xi(x) \exp \left(-{\bf{A}}x\right) {\bf{B}}
\exp \left( {\bf{A}} x \right) .
\label{ranmat}
\end{equation}
Van Kampen has established a set of systematic rules for constructing
all the terms of the series~(\ref{cumexp}); applying his prescriptions
one obtains that all generators of odd order are zero because the odd
moments of the potential vanish.
As for the even-order generators, the zero-th order term has the simple
form ${\bf{K}}_{0} = {\bf{A}}$, which corresponds to ignoring the
random potential.
The first term where disorder manifests its effect is the second-order
generator
\begin{displaymath}
{\bf{K}}_{2} (x) = \int_{0}^{x} dx_{1}
\exp \left( {\bf{A}} x \right)
\langle {\bf{M}}(x) {\bf{M}}(x_1) \rangle
\exp \left( -{\bf{A}} x \right) ,
\end{displaymath}
which is proportional to the two-point correlator; a refinement of this
result is obtained by taking into account the fourth-order term
\begin{displaymath}
\begin{array}{l}
{\bf{K}}_{4} (x) =
\int_{0}^{x} dx_1 \int_{0}^{x_1} dx_2 \int_{0}^{x_2} dx_3
\exp \left( -{\bf{A}} x \right) \\
\times \left[
\langle {\bf{M}}(x) {\bf{M}}(x_{1}) {\bf{M}}(x_{2}) {\bf{M}}(x_{3})
\rangle \right. \\
- \langle {\bf{M}}(x) {\bf{M}}(x_{1}) \rangle
\langle {\bf{M}}(x_{2}) {\bf{M}}(x_{3}) \rangle \\
- \langle {\bf{M}}(x) {\bf{M}}(x_{2}) \rangle
\langle {\bf{M}}(x_{1}) {\bf{M}}(x_{3}) \rangle \\
- \left. \langle {\bf{M}}(x) {\bf{M}}(x_{3}) \rangle
\langle {\bf{M}}(x_{1}) {\bf{M}}(x_{2}) \rangle \right]
\exp \left( {\bf{A}} x \right)
\end{array}
\end{displaymath}
whose integrand constitutes a non-trivial example of ordered cumulant
of the matrix~(\ref{ranmat}).
Although the rules of Van Kampen make possible to derive every partial
generator ${\bf{K}}_{n}(x)$, for the purpose of this work we can
truncate the series~(\ref{cumexp}) and neglect all terms of higher order
than the fourth.
As can be seen from Eq.~(\ref{nonstoc}), the behaviour of $\langle u(x)
\rangle$ for $x \rightarrow \infty$ is determined by the eigenvalue
with the largest real part of the asymptotic generator ${\bf{K}}(\infty)
= \lim_{x \rightarrow \infty} {\bf{K}} (x)$; we are thus led to the
conclusion that, in order to determine the Lyapunov
exponent~(\ref{loclen}) with fourth-order accuracy, we have to solve the
secular equation for the truncated asymptotic generator
\begin{equation}
\overline{{\bf{K}}}(\infty) = {\bf{A}} + \varepsilon^{2}
{\bf{K}}_{2}(\infty) + \varepsilon^{4} {\bf{K}}_{4}(\infty) .
\label{truncgen}
\end{equation}

Carrying out the calculations, one obtains that the inverse localisation
length~(\ref{loclen}) can be written as
\begin{equation}
\lambda (k) = \varepsilon^{2} \lambda_{2}(k) + \varepsilon^{4}
\left( \lambda_{4}^{(G)}(k) + \lambda_{4}^{(NG)}(k) \right) +
o\left( \varepsilon^{4} \right)
\label{lyap}
\end{equation}
where the second-order term has the standard form
\begin{equation}
\lambda_{2}(k) =
\frac{1}{4 k^{2}} \int_{0}^{\infty} \chi_{2} (x) \cos(2kx) dx .
\label{lambda2}
\end{equation}
As for the fourth-order term of the Lyapunov exponent~(\ref{lyap}), its
first component can be written as
\begin{equation}
\begin{array}{l}
\lambda_{4}^{(G)} (k) =
\frac{1}{16 k^{5}} \varphi_{s}(k,0) \left[ \varphi_{c}(k,0) +
4 \varphi_{c}(0,0) \right] \\
- \frac{1}{8 k^{4}} \left[ \varphi_{c}(k,0) + 2 \varphi_{c}(0,0) \right]
\int_{0}^{\infty} \chi_{2}(x) x \cos (2kx) dx \\
+ \frac{1}{8k^{4}} \varphi_{s}(k,0)
\int_{0}^{\infty} \chi_{2}(x) x \sin (2kx) dx \\
+ \frac{1}{4k^{4}} \int_{0}^{\infty} \left[ \varphi_{c}(0,x) \varphi_{c}(k,x)
- \varphi_{c}^{2}(0,x) \cos(2kx) \right] dx \\
\end{array}
\label{l4g}
\end{equation}
where the functions $\varphi_{c}(k,x)$ and $\varphi_{s}(k,x)$ are
defined as
\begin{displaymath}
\begin{array}{ccc}
\varphi_{c}(k,x)  & = &
\int_{x}^{\infty} \chi_{2}(y) \cos(2ky) dy \\
\varphi_{s}(k,x) & = &
\int_{x}^{\infty} \chi_{2}(y) \sin(2ky) dy \\
\end{array}.
\end{displaymath}
The second component turns out to be
\begin{equation}
\begin{array}{r}
\lambda_{4}^{(NG)} (k) = \frac{1}{k^{4}} \int_{0}^{\infty} dx
\int_{0}^{x} dy \int_{x}^{\infty} dz \Delta_{4} (y,x,z) \\
\times \left[ \cos(2ky) \cos(2kz-2kx) - \cos(2kz) \right] \\
\end{array}
\label{l4ng}
\end{equation}
where the symbol $\Delta_{4} (x_1,x_2,x_3)$ represents the fourth
cumulant of the random potential.
These formulae show that, in contrast to the relative simplicity of the
second-order result~(\ref{lambda2}), the fourth-order terms~(\ref{l4g})
and~(\ref{l4ng}) exhibit a non-trivial dependence on the binary correlator
and on the fourth cumulant, respectively.
The greater complexity of the results is a consequence of the more
sophisticated description of localisation, which is visualised as
an interference effect generated by double scatterings of the electron
in the second-order scheme, whereas the fourth-order approximation also
takes into account quadruple scattering processes.
The more accurate description of the localisation processes makes
possible to draw a distinction between Gaussian and non-Gaussian
disorders. These two classes of random potentials are necessarily
undifferentiated within the framework of the second-order approximation,
since the second-order Lyapunov exponent~(\ref{lambda2}) depends only
on the second moment of the potential; in the fourth-order description,
however, the differences between the two kinds of potentials emerge in
the component~(\ref{l4ng}) which, being a function of the cumulant,
vanishes if, and only if, the disorder is Gaussian.
In conclusion, the separation of the fourth-order term of the
Lyapunov exponent in the two components~(\ref{l4g}) and~(\ref{l4ng})
is justified by the distinct physical character of these two parts:
while the former is common to all Gaussian and non-Gaussian potentials
with the same two-point correlation function, the latter describes the
specific effect of the non-Gaussian nature of the disorder.

To analyse the case of disorder with long-range correlations, we
evaluated the form taken by expressions~(\ref{lambda2}) and~(\ref{l4g})
when the correlation function is
\begin{equation}
\chi_{2}(x) = \sigma_{2} \frac{\sin(2k_{c}x)}{x} \exp(-x/l_{c})
\label{frchi}
\end{equation}
and we subsequently took the limit $l_{c} \rightarrow \infty$, which
reduces the correlation function to the form
\begin{equation}
\chi_{2}(x) = \sigma_{2} \frac{\sin(2k_{c}x)}{x} .
\label{lrchi}
\end{equation}
This approach is generally successful, but it fails for $k=k_{c}$ because
for this value of the wavevector the asymptotic limit of the
generator~(\ref{cumexp}) exists only as long as $l_{c}$ is finite.
Our results for the Lyapunov exponent, therefore, are not
valid in a small neighbourhood of $k_{c}$.

\begin{figure}[htp]
\begin{center}
\caption{Fourth-order term $\lambda_{4}^{(G)}(k)/\sigma_{2}^{2}$ vs.
$k/k_{c}$}
\label{l4fig}
\epsfig{file=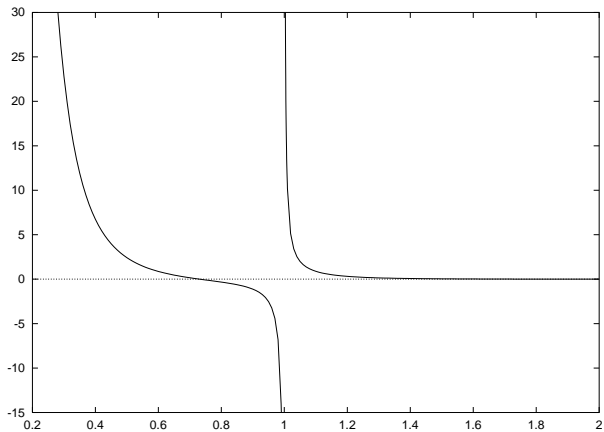,width=3.3in,height=2.3in}
\end{center}
\end{figure}

Inserting the correlation function~(\ref{frchi}) in Eq.~(\ref{lambda2})
and taking the limit $l_{c} \rightarrow \infty$, one obtains that
the second-order Lyapunov exponent has the form
\begin{equation}
\lambda_{2}(k) = \left\{ \begin{array}{ccl}
                  \frac{\sigma_{2} \pi}{8k^{2}} & \mbox{for} & 0<k<k_{c} \\
                                    0           & \mbox{for} &  k_{c}<k  \\
                          \end{array} \right. .
\label{lrl2}
\end{equation}
Eq.~(\ref{lrl2}) shows that long-range correlations of the
form~(\ref{lrchi}) make the second-order Lyapunov exponent vanish when
the wavevector $k$ exceeds a critical value $k_{c}$. This result was
interpreted in~\cite{Izr99} as evidence that long-range correlation of
the potential can create a continuum of extended states, with
a mobility edge at $k = k_{c}$.
Second-order theory, however, cannot precise the nature of the
``extended'' states nor define their spatial extension. To overcome
this limit, one can compute the fourth-order term~(\ref{l4g}) of the
Lyapunov exponent.
Evaluating expression~(\ref{l4g}) for the correlation function~(\ref{frchi})
and then taking the limit $l_{c} \rightarrow \infty$, one arrives at
the result
\begin{equation}
\lambda_{4}^{(G)}(k) = \left\{
\begin{array}{lcl}
\Lambda_{1}(k) & \mbox{for} & 0 < k < k_{c} \\
\Lambda_{2}(k) & \mbox{for} & k_{c} < k < 2k_{c} \\
0  & \mbox{for} & 2 k_{c} < k \\
\end{array} \right.
\label{lrl4}
\end{equation}
where the functions $\Lambda_{1}(k)$ and $\Lambda_{2}(k)$ are defined
as
\begin{displaymath}
\begin{array}{ccl}
\Lambda_{1}(k) & = & \frac{\sigma_{2}^{2}}{4k^{4}} \left[ \frac{5\pi}{32k}
\ln \left( \frac{k+k_{c}}{k-k_{c}} \right)^{2}
- \frac{\pi}{8} \frac{k^{2}+kk_{c}+k_{c}^{2}}{k_{c}(k_{c}^{2}
-k^{2})} - I(k) \right] \\
\Lambda_{2}(k) & = & \frac{\sigma_{2}^{2}}{4k^{4}} \left[\frac{\pi}{8k}
\ln \left( \frac{k+k_{c}}{k-k_{c}} \right)^{2}
+ \frac{\pi}{8} \frac{2k_{c}-k}{k_{c}(k-k_{c})} - I(k) \right]
\end{array}
\end{displaymath}
with
\begin{displaymath}
I(k) = \int_{0}^{\infty} \mbox{si}^{2}(2k_{c}x) \cos(2kx) dx .
\end{displaymath}
The symbol $\mbox{si}(x)$ in the previous equation represents the sine
integral defined as $\mbox{si}(x) = -\int_{x}^{\infty} dt \sin(t)/t$.
In Fig.~\ref{l4fig} we show the plot of the fourth-order Lyapunov
exponent~(\ref{lrl4}) as a function of the wavevector $k$.
As already observed, the divergence for $k = k_{c}$ must be disregarded
because the asymptotic generator~(\ref{truncgen}) is not defined for
this value of the wavevector in the limit of long-range correlations.
In the interval $0<k<k_{c}$ the second-order Lyapunov exponent~(\ref{lrl2})
is positive and the fourth-order term~(\ref{lrl4}) constitutes just a small
correction which enhances or decreases the localisation of electronic
states according to whether it assumes positive or negative values.
The main interest of Eq.~(\ref{lrl4}), however, lies in the fact that
it shows that the fourth-order Lyapunov exponent is positive in the
interval $k_{c}<k<2k_{c}$, where the second-order term~(\ref{lrl2})
vanishes.
This result clarifies the nature of the mobility edge at $k=k_{c}$:
the electronic states are exponentially localised on both sides of the
threshold, but the localisation length is increased by a factor
$O(1/\varepsilon^{2})$ when the wavevector exceeds the critical value
$k_{c}$. For weak disorder, the corresponding change in the electron
mobility is so conspicuous that one can legitimately speak of
``delocalisation transition'', even if the ``extended'' states are not
completely delocalised like Bloch states in a crystal lattice.
A second noteworthy aspect of the fourth-order term~(\ref{lrl4}) is that
it vanishes for $k > 2k_{c}$ so that, in the case of Gaussian disorder,
a second transition takes place at $k = 2k_{c}$ with the electrons becoming
even more delocalised.

So far our discussion has been centred on the term~(\ref{l4g}), which
allows a complete analysis of Gaussian potentials.
For non-Gaussian disorder, however, one must also take into account the
term~(\ref{l4ng}) which enhances localisation.
As an example, one can consider the case of a non-Gaussian potential with
long-range binary correlations of the form~(\ref{lrchi}) and with an
exponentially decaying fourth cumulant
\begin{equation}
\Delta_{4} (x,y,z) = \sigma_{4} \exp \left[ - \beta \left( |x|
+ |y| + |z| \right) \right] .
\label{expcum}
\end{equation}
In this case the fourth-order term of the Lyapunov exponent is the
sum of the component~(\ref{lrl4}) and of the additional term
\begin{equation}
\lambda_{4}^{(NG)}(k) = \frac{15\sigma_{4}\beta}{8k^{2}
(\beta^{2} + k^{2}) (\beta^{2} + 4k^{2}) (9\beta^{2} + 4k^{2})}
\label{expl4ng}
\end{equation}
which is obtained by inserting expression~(\ref{expcum}) in Eq.~(\ref{l4g}).
Since the extra term~(\ref{expl4ng}) is always positive, all electronic
states with $k>k_{c}$ are exponentially localised with inverse localisation
length proportional to the fourth power of the disorder strength,
$\lambda(k) \propto \varepsilon^{4}$.

The main conclusion of this study is that the delocalisation transition
occurring in 1D models with long-range correlated disorder consists
in a change of the scaling behaviour of the Lyapunov exponent when the
electronic energy crosses a critical threshold: more specifically, the
Lyapunov exponent, which is a {\em quadratic} function of the disorder
strength in the localised phase, assumes a {\em quartic} form in the
extended phase.
Another relevant remark is that non-Gaussian potentials produce a
stronger localisation effect than their Gaussian counterparts.

The author wishes to thank S. Ruffo and A. Politi for useful discussions
and is grateful to F. M. Izrailev for valuable comments and suggestions.

\end{document}